\documentclass[sigconf]{acmart}

\usepackage{amsmath}
\AtBeginDocument{%
  }
\usepackage{listings}
\usepackage[linesnumbered, ruled, vlined]{algorithm2e}

\copyrightyear{2026}
\acmYear{2026}
\setcopyright{cc}
\setcctype{by}
\acmConference[]{}
\acmBooktitle{}
\acmConference[SCA/HPCAsiaWS 2026]{SCA/HPCAsia 2026 Workshops: Supercomputing Asia and International Conference on High Performance Computing in Asia Pacific Region Workshops}{January 26--29, 2026}{Osaka, Japan}
\acmPrice{}
\acmDOI{}
\acmISBN{}

\begin{document}

\def\thankses{%
This is the author's version of the work. It is posted here for your personal use. 
The definitive version was published in 
SCA/HPCAsiaWS '26: Proceedings of the Supercomputing Asia and International Conference on High Performance Computing in Asia Pacific Region Workshops,
\url{https://doi.org/10.1145/3784828.3785398}.
}

\title{Mixed precision solvers with half-precision floating point numbers for Lattice QCD on A64FX processor}

\author{Issaku Kanamori}
\email{kanamori-i@riken.jp}
\orcid{0000-0003-4467-1052}
\affiliation{%
  \institution{RIKEN Center for Computational Science}
  \city{Kobe}
  \country{Japan}
}

\author{Hideo Matsufuru}
\email{hideo.matsufuru@kek.jp}
\orcid{0000-0003-1056-3969}
\affiliation{%
  \institution{Computing Research Center,
  High Energy Accelerator Research Organization (KEK)
  and Accelerator Science Program,
  Graduate Institute for Advanced Studies, SOKENDAI}
    \city{Tsukuba}
  \country{Japan}
}

\author{Tatsumi Aoyama}
\email{aoym@issp.u-tokyo.ac.jp}
\orcid{0009-0009-0491-1024}
\affiliation{%
\institution{The Institute for Solid State Physics, The University of Tokyo}
\city{Kashiwa}
\country{Japan}
}

\author{Kazuyuki Kanaya}
\email{kanaya@ccs.tsukuba.ac.jp}
\orcid{0000-0002-5608-6347}
\affiliation{%
\institution{Tomonaga Center for the History of the Universe, University of Tsukuba}
\state{Tsukuba}
\country{Japan}
}

\author{Yusuke Namekawa}
\email{namekawa@fukuyama-u.ac.jp}
\orcid{0000-0002-3578-5085}
\affiliation{%
  \institution{Department of Computer Science, Fukuyama University}
 \city{Fukuyama}
\country{Japan}
}

\author{Hidekatsu Nemura}
\email{hidekatsu.nemura@rcnp.osaka-u.ac.jp}
\orcid{0009-0004-8203-4569}
\affiliation{%
\institution{Research Center for Nuclear Physics, Osaka University}
\city{Ibaraki-shi}
\country{Japan}
}

\author{Keigo Nitadori}
\email{keigo@riken.jp}
\orcid{0000-0001-7374-4236}
\affiliation{%
  \institution{RIKEN Center for Computational Science}
  \city{Kobe}
  \country{Japan}
}
\renewcommand{\shortauthors}{Kanamori et al.}

\begin{abstract}
We investigate the use of half-precision floating-point numbers (FP16) in mixed-precision linear solvers for lattice QCD simulations.
Since the emergence of GPUs for general-purpose,  mixed-precision algorithms that combine single-precision (FP32) with double-precision (FP64) arithmetics have become widely used in this field and others.
While FP32-based methods are now well established, we examine the practicality of using FP16.
In this work, we introduce rescaling steps in
both the outer iterative refinement step and the inner BiCGStab solver
to avoid numerical instability.
In our experiments with a simple Wilson kernel, the solver shows improved stability, and the additional iteration count compared to the FP64 version remains within 20\%, indicating that the FP16 version is practical for use.
We believe that the proposed rescaling methods can also benefit other mixed precision preconditioners
in avoiding underflows.
\end{abstract}



\maketitle

\section{Introduction}
\label{sec:introduction}

Most scientific computations require FP64 (64-bit floating-point)
arithmetic precision in their results.
Lower precisions, typically FP32 arithmetics, offer advantages
in the performance of floating-point number operations,
due to their smaller amount of data loaded from the memory per
arithmetic operation.
Furthermore, there are architectures, particularly those equipped with SIMD
arithmetic operation units, 
that can simultaneously process
a larger number of data with a shorter bit length.
Thus, lower precision arithmetics have been utilized in parts of
computations in which less precision is sufficient.
A typical example is the lower precision preconditioning in iterative
solvers for linear equation systems
\cite{buttari:hal-05071696, Lindquist9462418, Higham_Mary_2022}.

Recently, the increasing demand for artificial intelligence (AI) applications
has changed the trend of processors.  
As low-precision arithmetics are sufficient for the AI, 
processors tend to possess more powerful arithmetic operation units
for the FP16 (half-precision) operations specified in IEEE 754.
In particular, GPUs have made use of FP16 for graphics computation
and have been optimized for AI applications over the last decade.
For example, NVIDIA equipped the Tensor Core units in its Volta generation
and has continuously enhanced FP16 performance.
Since RIKEN has decided to adopt NVIDIA GPUs as accelerators of the Japanese next
generation national flagship supercomputer, known as Fugaku NEXT,
exploiting the high performance of FP16 with Tensor Core is also essential
for scientific applications.
The current flagship system, the Supercomputer Fugaku, also supports FP16
operations in its genuine SIMD arithmetic units on the A64FX
processor architecture \cite{Matsuka2020_9355239}.
Therefore, an application developed for the A64FX processor can be adopted
for FP16 at least from an algorithmic standpoint.
This is important not only as preparation for the future large-scale system,
but also for accelerating simulations on Fugaku.

In this work, we focus on the A64FX processor and examine the applicability
of FP16 arithmetics to large-scale scientific simulations.
The A64FX instructions support the scalable vector extension (SVE)
with a 512-bit SIMD vector enabling four times higher performance
of the FP16 arithmetics compared to FP64.
As the target application, we adopt lattice QCD,
a discretized theory of QCD (Quantum Chromodynamics) that describes
the dynamics of elementary particles known as quarks and gluons.
Lattice QCD is defined on a four-dimensional lattice, and its
quantum expectation values can be determined by Monte Carlo simulations \cite{Wilson_PhysRevD.10.2445}.
The most time-consuming part of the lattice QCD simulations is typically
an iterative solver for a discretized partial differential equation \cite{Finkenrath:2023sjg}.
For this linear equation solver, a mixed-precision preconditioning
using the FP32 arithmetics has been applied as a pioneering work to
exploit GPUs in the scientific numerical simulations \cite{Egri:2006zm}.

With the increasing importance of lower-precision arithmetics,
we apply the FP16 arithmetics to the mixed-precision solver in
lattice QCD.
More explicitly, we focus on the Wilson fermion matrix,
which is the simplest formulation of discretized quark actions.
Although in up-to-date large-scale simulations, its improved version,
known as the clover fermion matrix, is more widely used,
the simple structure of the Wilson fermion matrix is suitable for
an exploratory study in this paper.
The Wilson matrix is also used as a building block for another 
involved formulation known as the domain-wall fermion matrix,
which better preserves the important symmetry of QCD
compared to the other forms of fermion matrix, 
although it is computationally more demanding.
Thus, the investigation of the Wilson matrix still makes
practical sense.
As we report in this paper, the same algorithm used with FP32 does not work efficiently with FP16.
The proposed method in this paper to stabilize the iterative algorithm differs from those used
by other LQCD libraries. Moreover, the new algorithm may be beneficial for other applications
with low-precision solvers.

This paper is organized as follows.
In the next section, we briefly explain the lattice QCD and the
Wilson fermion matrix in some detail.
Section~\ref{sec:Algorithm} describes the mixed-precision
preconditioning in the iterative solver used in this work.
In Sec.~\ref{sec:Related_works}, the related works are summarized.
We explain the implementation details in Sec.~\ref{sec:Implementation}
and provide our results in Sec.~\ref{sec:Results}.
The final section is devoted to our conclusions and future outlooks.

\section{Lattice QCD and Wilson fermion matrix}
\label{sec:LQCD}

\begin{figure}[tb]
\noindent\centering%
\includegraphics[width=0.6\linewidth]{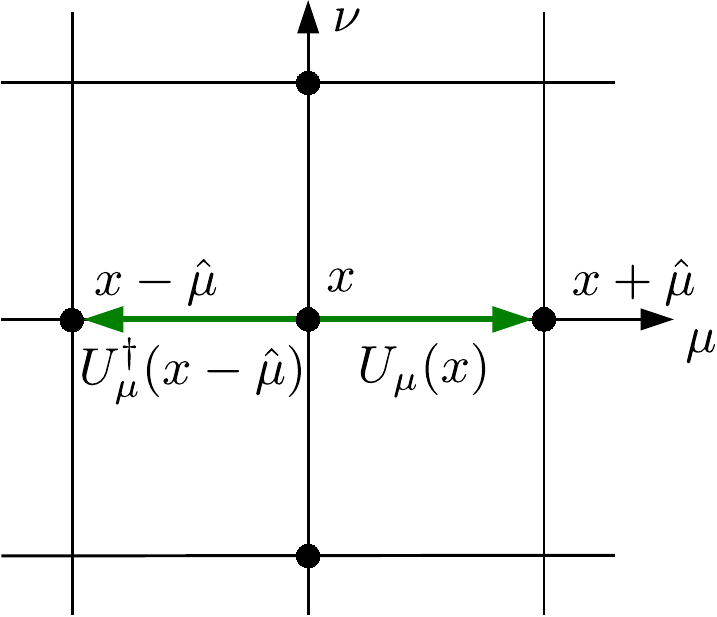}
\caption{
Two-dimensional sketch of the Wilson kernel. The quark fields $\psi(x)$
hops form $x$ to the neighboring lattice sites.
}
\label{fig:wilson}
\end{figure}

The strong interaction among quarks, one of the fundamental interactions
in the universe, is described by a theory called Quantum Chromodynamics (QCD).
QCD is a gauge field theory in which the interaction is mediated by
a gauge field called gluons.  We refer to the review sections in \cite{Particle_Data_PhysRevD.110.030001} by Particle Data Group for QCD itself.
In lattice QCD, which is a discretized formulation of QCD on
a four-dimensional Euclidean lattice, quarks and gluons are defined
on the lattice sites and links, respectively\footnote{
There are lots of textbooks and lecture notes on lattice QCD, including \cite{DeGrand_doi:10.1142/6065, Gattringer:2010zz, LesHouches_10.1093/acprof:oso/9780199691609.001.0001}. 
Ref.~\cite{Particle_Data_PhysRevD.110.030001} also contains a review of lattice QCD.  For an alternative approach to QCD other than lattice, see \cite{Magnifico:2024eiy}, for example. 
}.
Figure~\ref{fig:wilson} shows a two-dimensional sketch of the
lattice surrounding a site $x$ in the $\mu$-$\nu$ plane.
The gluon field $U_\mu(x)$ is placed on the link between the sites $x$ and
$x+\hat{\mu}$, where $\hat{\mu}$ is a unit vector in the $\mu$-direction.
We set the lattice spacing $a=1$ for simplicity.
Reflecting the number of color degrees of freedom, $N_c=3$, the gluon
field $U_\mu(x)$ represents a $3\times 3$ complex matrix 
and an element of the $\mathrm{SU}(3)$ group.
As $\mathrm{SU}(3)$ is a special unitary group, it follows that
$U_\mu(x)^\dagger = U_\mu(x)^{-1}$, {\it i.e.},
$(U_\mu(x))_{ab}^* = (U_\mu(x)^{-1})_{ba}$ for the matrix elements.
Multiplying $U_\mu(x)$ is a parallel translation and
its unitarity ensures that retracing back the same path returns to the original state.
The gluon field in the $(-\mu)$-direction at $x$ is therefore
$U^\dagger_\mu(x-\hat{\mu})$ as displayed in Fig.~\ref{fig:wilson}.

The quark field is theoretically defined as an anti-commuting
Grassmann field on sites, and it is not directly tractable on computers.
However, by integrating out the Grassmann field beforehand,
the quark field
can be represented as a vector on sites that carries the color and spinor
degrees of freedom.
This means that the vector on a site is an $N_c\times N_d = 3\times 4$
complex variables.
Here $N_d=4$ is the size of a Dirac spinor representing
the degrees of freedom for quark and anti-quark with
up and down spins.

The contribution of the quark field to the partition function becomes
$\exp[-\phi^\dagger (D^\dagger D)^{-1} \phi]$,
in the case of two degenerate dynamical flavors, where $D$ is
the fermion matrix mentioned in the introduction.
Since the condition of the lattice fermion matrix is that it approaches
the fermion operator of QCD in the continuum limit,
$a\rightarrow 0$, there is an arbitrariness in the definition of
the fermion matrix at finite $a$.
The aforementioned Wilson matrix is one of the simplest formulations
and is represented as%
\footnote{
A matrix with a different normalization, $D_W$ multiplied by $(4+M)$,
where $M=1/(2\kappa)-4$, is an equivalent representation and also commonly
used in the literature.
}
\begin{align}
(D_{\mathrm{W}})_{i a, j b}(x,y)
&= \delta_{i,j}\delta_{a,b}\delta_{x,y} 
 {}- \kappa \sum_{\mu=1}^4 \bigl[
  (1-\gamma_\mu)_{ij} (U_\mu)_{ab}(x)\delta_{x+\hat{\mu},y}
  \nonumber\\
& \qquad\qquad {} + (1+\gamma_\mu)_{ij} (U_\mu)_{ba}^*(x-\hat{\mu})\delta_{x-\hat{\mu},y}
\bigr],
\label{eq:Wilson_fermion_matrix}
\end{align}
where all component labels are explicitly indicated:
$a,b=1,2,3$ for color, $i,j=1,\dots, 4$ for spinor, and $x,y$ for
lattice site.
The so-called hopping parameter $\kappa$ corresponds to the quark mass $M$ through
the relation $\kappa=1/2(M+4)$,
although this relation is affected by dynamical effects such that
the massless limit shifts from $\kappa=1/8$.
$\gamma_\mu$ is a $4 \times 4$ matrix acting on the spinor space,
and the gluon field $U_\mu(x)$ is a $3\times 3$ complex matrix acting
on the color degree of freedom of a quark vector.
As depicted in Figure~\ref{fig:wilson}, the Wilson fermion matrix
contains only the nearest-neighbor coupling.
Thus, the matrix $D_W$ is a sparse complex matrix of rank 
$N_c N_d N_x N_y N_z N_t$, where $N_\mu$ ($\mu=x,y,z,t$) represents the lattice
extent in the $\mu$-direction.

\begin{algorithm}[tb]
\caption{
Pseudo code of a serial version of the Wilson fermion matrix operation:
$\psi = D_{\mathrm{W}} \phi$.
}
\label{alg:Wilson_code}
\For{$x$ in sites}{
 $\psi(x)=0$ \\
\For{$\mu$ in $x$-, $y$-, $z$-, $t$- directions}{
  $\phi^{(+)}(x+\hat{\mu}) ={}$ two-component spinor made from $\allowbreak{(1+\gamma_\mu)\phi(x+\hat{\mu})} $\\
  multiply $3\times 3$ link variables: $\phi^{(+)}(x) := U_\mu(x) \phi^{(+)}(x+\hat{\mu})$\\
  reconstruct four-component spinor from $\phi^{(+)}(x)$ and accumulate to $\psi(x)$\\
  $\phi^{(-)}(x-\hat{\mu}) = {}$ two-component spinor made from 
  $\allowbreak{(1-\gamma_\mu)\phi(x-\hat{\mu})} $ \\
  multiply $3\times 3$ link variables: ${\phi^{(-)}(x) := 
  U_\mu^\dagger(x-\hat{\mu}) \phi^{(-)}(x-\hat{\mu})}$\\
  reconstruct four-component spinor from $\phi^{(-)}(x)$ and accumulate to $\psi(x)$
}
}
\end{algorithm}

Let us examine the operations in the matrix $D_W$ more closely.
It represents a nine-point stencil computation, accumulating data from the eight
nearest neighboring sites.
Since the number of arithmetic operations is important in low-precision 
contexts, we explain the computational steps
corresponding to the
second term in the right-hand side of Eq.(\ref{eq:Wilson_fermion_matrix})
(prior to the multiplying by $-\kappa$) in Algorithm~\ref{alg:Wilson_code}.
Let us consider the operation in the $x$-direction within the second for loop.
(Note that we reused the letter $x$ for both direction and site index,
$a$ for lattice spacing and color index.)
First, the vector at the neighboring site, $\phi(x+\hat{\mu})$,
is loaded.
Multiplying by $\gamma_\mu$ 
permutes the spinor components.
Explicitly, with our convention of the $\gamma_\mu$ matrices,
it is
\begin{align}
\left(\begin{array}{c}
 \phi^{(+)}_{1} \\
 \phi^{(+)}_{2} \\
 \phi^{(+)}_{3} \\
 \phi^{(+)}_{4} \\
\end{array} \right)
&= 
\left( 1 - \gamma_1\right)
\left(\begin{array}{c}
 \phi_{1} \\
 \phi_{2} \\
 \phi_{3} \\
 \phi_{4} \\
\end{array} \right)
= 
\left(\begin{array}{cccc}
  1 &  0 & 0  & i \\
  0 &  1 &  i & 0 \\
  0 & -i &  1 & 0 \\
 -i &  0 &  0 & 1 \\
\end{array} \right)
\left(\begin{array}{c}
 \phi_{1} \\
 \phi_{2} \\©
 \phi_{3} \\
 \phi_{4} \\
\end{array} \right)
\nonumber\\
&= 
\left(\begin{array}{c}
 \phi_{1} + i \phi_{4} \\
 \phi_{2} + i \phi_{3} \\
 -i(\phi_{2} + i\phi_3) \\
 -i(\phi_{1} + i\phi_4) \\
\end{array} \right) ,
\end{align}
where we omit color and site indices.
For the identities 
$\phi^{(+)}_{3} =-i \phi^{(+)}_{2}$
and
$\phi^{(+)}_{4} =-i \phi^{(+)}_{1}$,
only two out of the four components, $\phi^{(+)}_{1}$ and $\phi^{(+)}_{2}$ 
need to be multiplied by the $3\times 3$ matrix $U_1(x)$.
The resulting products are accumulated into $\psi(x)$.
Similar operations are conducted for the $-x$, $\pm y$, $\pm z$, and
$\pm z$-directions.
Because of the efficiency detailed later, the implementation with this decomposition
is widely used in the practical lattice QCD simulation codes.

In the end, the hopping operation between a nearest-neighbor lattice site  
is accomplished by the following three steps.
\begin{enumerate}
    \item Four-to-two transformations:
    For each color, a projection operator $(1\pm\gamma_\mu)$ is multiplied to a Dirac spinor, effectively retaining two out of the four elements.
    For three colors, this results in a total of six complex numbers. 
    \item Three-to-three transformations:
    For each two-spinor, the color transformation is executed using
    the $\mathrm{SU}(3)$ matrix $U$ or $U^\dag$.
    \item Two-to-four reconstructions:
    The two omitted components of the four-spinor are reconstructed by simple permutation and sign modification operations, accumulating twelve complex numbers.
\end{enumerate}

Formally, this hopping operation can be 
achieved by multiplying a complex $12 \times 12$ block matrix.
Such representation allows the use of general-purpose sparse linear algebra libraries,
provided they support block matrices as elements.
However, this approach sacrifices efficiency:
it neglects the fact that the transformation on the color and spinor indices can be
factored out, and the fact that the effective degrees of freedom for the spinor are half of the full system.
Expressing such structures in the general-purpose libraries is difficult.
In terms of floating-point number operations,
implementation with the above decomposition requires
168 FLOPs, while a multiplication of a $12\times 12$ complex matrix on a vector requires
$12\times 12 \times 8 = 1152$ FLOPs.
The overhead is sixteen times in the memory footprint and bandwidth to hold
a $12 \times 12$ matrix instead of $3 \times 3$ for each hopping, which is usually
unacceptable for memory-bound applications.

The equation to be solved is
\begin{equation}
 \sum_{k} D_{\mathrm{W}}(j,k) \xi (k) = \eta(j)
 \label{eq:linear_equation},
\end{equation}
where $\eta$ is a source vector, and
$j$ and $k$ are collective indices of color, spinor, and site
degrees of freedom.
The even-odd preconditioning is frequently adopted
to accelerate solving the above linear equation
(\ref{eq:linear_equation}).
Dividing the lattice sites into even and odd sites, 
Eq.~(\ref{eq:linear_equation}) is equivalent to 
\begin{equation}
    D_\mathrm{W} \xi
    =
    \begin{pmatrix}
      D_{\mathrm{ee}} & D_{\mathrm{eo}} \\
      D_{\mathrm{oe}} & D_{\mathrm{oo}} \\
    \end{pmatrix}
    \begin{pmatrix}
      \xi_{\mathrm e} \\
      \xi_{\mathrm o}
    \end{pmatrix}
    =
    \begin{pmatrix}
      \eta_{\mathrm e} \\
      \eta_{\mathrm o}
    \end{pmatrix}
    = \eta.
\end{equation}
Now $\xi_{\mathrm e}$ can be solved independently as in
\begin{equation}
 \left[ 1-D_{\mathrm{ee}}^{-1} \, D_{\mathrm{eo}} \, D_{\mathrm{oo}}^{-1} \, D_{\mathrm{oe}}\right] \xi_{\mathrm{e}}
 = D_{\mathrm{ee}}^{-1} \left(\eta_{\mathrm{e}}
  - D_{\mathrm{eo}} \, D_{\mathrm{oo}}^{-1} \, \eta_{\mathrm{o}}\right) ,
 \label{eq:even-odd_preconditioned_equation}
\end{equation}
where a new matrix to be solved is defined in the square bracket,
and a new known vector as the right-hand side.
Once $\xi_{\mathrm e}$ is determined, $\xi_{\mathrm o}$ is obtained as
\begin{equation}
 \xi_{\mathrm{o}} = D_{\mathrm{oo}}^{-1} \left(\eta_{\mathrm{o}}
 - D_{\mathrm{oe}}\, \xi_{\mathrm{e}}\right).
\end{equation}
In the case of the Wilson fermion matrix, the diagonal blocks
$D_{\mathrm{ee}}$ and $D_{\mathrm{oo}}$ are simply unit matrices,
and their inverses are trivial.
The arithmetic operations of $D_{\mathrm{eo}}$ and $D_{\mathrm{oe}}$ are
eight-point stencil operations, which are
in total the same as those of the original $D_{\mathrm{W}}$
except for the self site.
However, the operator on the left-hand side of
Eq.~(\ref{eq:even-odd_preconditioned_equation})
generally has
a smaller condition number than the original
matrix $D_{\mathrm{W}}$, and thus it accelerates the convergence.
In addition, the working vectors in the iterative linear solver
require only  half the size of full vectors, 
offering an advantage in memory usage.

\section{Algorithm for the Mixed Precision Solver}
\label{sec:Algorithm}

\begin{algorithm}[tb]
\caption{
Preconditioned Richardson algorithm.\\
Datatype conversions are denoted as C++ like operators, double() and low\_prec().
}
\label{alg:mixed_org}
$x=0$,\ $r=b$ \\
  \While{$|r|$ is not small enough}{
  $s=|r|$, $y=r/s$\\
  solve low precision system: $\tilde{A} \tilde{t} = \text{low\_prec}(y)$ for $\tilde{t}$\\
  $x:= x + \frac{1}{s} \text{double}(\tilde t)$\\
  $r:= {b} -A x$\\
  }
\end{algorithm}

\begin{algorithm}[tb]
\caption{
An improved version of iterative refinement with rescaling to solve $Ax=b$.  
Here, $\tilde{A}$ is a low-precision approximation of the matrix $A$,
for which we assume FP16.
It also contains a recalculation of the scaling factor.
The recalculated factor, $\hat{s}$, is calculated in FP32 together with a global
reduction in FP64 (see Fig.~\ref{fig:norm}). 
}
\label{alg:mixed}
$x=0$,\ $r=b$,\ $s=\text{given}$ \\
  \While{$|r|$ is not small enough}{
  rescale the residual vector: $y = \alpha r$\ s.t.\ $|y|=s$\\
  $\tilde{y}=\text{low\_prec}(y)$ \\
  $\hat{s} = |\text{single}(\tilde{y})|$ \\  
 \label{line:hat_s}
  solve low precision system: $\tilde{A} \tilde{t} = \tilde{y}$ for $\tilde{t}$\\
  $x:= x + \frac{1}{\hat{s}} \text{double}(\tilde t)$\\
  $r:= {b} -A x$\\
  }
\end{algorithm}

We solve the linear equation (\ref{eq:even-odd_preconditioned_equation})
using an iterative refinement algorithm that incorporates a low-precision preconditioning.
Algorithm~\ref{alg:mixed_org} illustrates the Richardson algorithm with preconditioning,
which we employ in this work.
Solving the low-precision linear equation,
$\tilde{A} \tilde{t} = \tilde{y}=\text{low\_prec}(y)$, plays the role
of preconditioning, for which any iterative algorithm can be employed.
Since the matrix we employ, as defined in
Eq.~(\ref{eq:even-odd_preconditioned_equation}), is complex and
non-Hermitian, we adopt the BiCGStab algorithm \cite{van_der_Vorst_10.1137/0913035}.
Although we do not use it in this work,
a flexible variant is also know \cite{VOGEL2007226}, which can be combined with low-precision preconditioners.

In the later section, however, we will find that applying
Algorithm~\ref{alg:mixed_org} as-is is inefficient for FP16 preconditioning. 
In our understanding, this is due to the limited range of the exponent in FP16 numbers.
Since the norm of the residual vector decreases during iteration,
we scale the residual vector by introducing a normalization factor
in order for more efficient use of the range of the exponent.
Thus, we also examine the rescaled version of the iterative refinement,
as displayed in Algorithm~\ref{alg:mixed}, together with the
rescaled BiCGStab algorithm, Algorithm~\ref{alg:bicgstab_rescaled}.
Note that setting the rescaling parameter $\gamma$ to $\gamma=1$
throughout the iteration corresponds to the original BiCGStab
algorithm without rescaling.
The algorithm in Algorithm~\ref{alg:bicgstab_rescaled} also
incorporates the improved prescription proposed in
\cite{BiCGStab_with_omega}, for which the parameter $\omega_0$
is introduced.
We set $\omega_0=0.7$ in this work.
In practice, the initial guess of the solution, $x_0$, is set
to $x_0=0$ in this work.
These improvements aim to avoid underflow in the residual vector
of the BiCGStab algorithm.
As the iteration steps proceed inside the low-precision BiCGStab solver,
the residual norm becomes smaller.
The elements of the residual vector accordingly become smaller,
and some of them may underflow during iterations.
By rescaling the residual norm, or by having a large input vector from
the beginning, one can expect fewer occurrences of underflow.

We also introduce an additional trick to reduce the effect of rounding error
in these algorithms.
Due to the rounding error, the normalization of the rescaled FP16 vectors may
deviate significantly from the intended normalization.
In Algorithm~\ref{alg:mixed}, this is corrected by using $\hat{s}$
instead of $s$ at line \ref{line:hat_s}.  Here, we assume that the calculation of $\hat{s}$ is
accurate enough and does not spoil the refinement processes.
A similar trick is introduced at line \ref{line:recaclulate_gamma} in
Algorithm~\ref{alg:bicgstab_rescaled}, where $\gamma$ is recalculated
by taking a ratio of the residual norms before and after rescaling.

The initial value of $\rho'$ in Algorithm \ref{alg:bicgstab_rescaled} may
require some explanation.
As the initial probe vectors $p$ and $v$ are set to zero, the result is
mathematically independent of the choice of the initial value of $\rho'$
and $p=r$ after the update during the first iteration (but before the rescaling).
This is numerically not always correct.
If the value of $\beta$, which is treated as an FP64 variable,
exceeds the maximum number of FP16, then the probe vector $p$
becomes $\pm\infty$, subsequently leading to NaN in the later operations.
If the initial $\rho'$ is $1$, for example,
as in the original implementation in Bridge++ for FP64 and FP32,
this in fact occurs if the norm of the input vector is large.  The value in algorithm \ref{alg:bicgstab_rescaled}
is chosen to give $\beta=1$ in the first iteration.

In Algorithm \ref{alg:bicgstab_rescaled2}, we introduce a new factor $\lambda$ 
to rescale the solution vector in addition to the residual vector. 
In contrast to avoiding underflow of the residual vector by $\gamma$, 
this factor $\lambda$ aims to avoid an overflow.
If the system we want to solve has one or more very small eigenvalues,
and the input vector has a large overlap with such eigenmodes, the norm of
the solution can be much larger than the norm of the input vector and
may overflow in FP16.
The rescaled solution can also help keep the magnitude of working vectors and
the solution vector in the same range.
Although one can also introduce recalculation of the rescaling factor $\lambda$
as introduced to $\gamma$ in line~\ref{line:recaclulate_gamma} in
Algorithm~\ref{alg:bicgstab_rescaled}, we adopt this prescription for simplicity
in this work.

\begin{algorithm}[tb]
\caption{
BiCGStab solver to solve $Ax=b$ with rescaling.
It also contains an improving prescription proposed in \cite{BiCGStab_with_omega}.
In practice, we use $x_0=0$ and $\omega_0=0.7$ in this work.
By setting $\gamma=1$ at every step, one can recover
the standard BiCGStab without rescaling.
}
\label{alg:bicgstab_rescaled}
\newcommand{\mycomment}[1]{\hfill\makebox[0.4\linewidth][l]{\textit{#1}}}
 $x=x_0$, $r=b-Ax$, $\sigma=\text{given}$ \\
  $r^*=r$ , $p=0$, $v=0$, $\rho'=|r|^2$ \\
  $\omega'=\alpha'=\gamma'=1$\\
 \While{$|r|/\gamma'$ is not small enough}{
 $\rho=(r^*,r)$, $\beta= \frac{\rho \alpha'}{\rho'\omega'}$\\
 $p:= \beta ( p-\omega' v ) + r$ \mycomment{// probe vector}\\
 $v= Ap$ \\
 $\alpha = \frac{\rho}{(r^*, v)}$ \\
 $r:=r-\alpha v$ \\
 $t= Ar$ \\
 $\omega = \frac{(t,r)}{(t,t)}$ , $\hat{\omega} = \frac{|(t,r)|}{\sqrt{ (t,t) (r,r)}}$\\
 if($ \hat{\omega}  < \omega_0$) {
 $\omega:= \omega \omega_0/\hat{\omega}$
 }\\
 $x:= x+ \frac{1}{\gamma'} \omega r + \frac{1}{\gamma'}\alpha p$ \mycomment{// update the solution}\\
 $r:= r -\omega t$  \mycomment{// update the residual vector} \\
 $g=|r|$, $\gamma = \sigma /g$\\
 $r:= \gamma r$ \\
 $\gamma:= g/|r|$ \mycomment{// recalculate $\gamma$} \\
 \label{line:recaclulate_gamma}
 $p:=\gamma p$, $v:=\gamma v$ \mycomment{// rescaling working vectors}\\
 $\omega'=\omega$, $\alpha'=\alpha$, $\rho'=\gamma \rho$\\
 $\gamma':=\gamma' \gamma$
}
\end{algorithm}

\begin{algorithm}[tb]
\caption{
A variant of BiCGStab solver to solve $Ax=b$ with rescaling,
which also rescales the solution vector.
In practice, we set $\sigma_\lambda=\sigma$.
By setting $\lambda=\lambda'=1$ at every step,
Algorithm.~\ref{alg:bicgstab_rescaled} is recovered.
}
\label{alg:bicgstab_rescaled2}
\newcommand{\mycomment}[1]{\hfill\makebox[0.4\linewidth][l]{\textit{#1}}}
 $x=x_0$, $r=b-Ax$, $\sigma=\text{given}$, $\sigma_\lambda=\text{given}$ \\
  $r^*=r$ , $p=0$, $v=0$, $\rho'=|r|^2$ \\
  $\omega'=\alpha'=\gamma'=1$\\
 \While{$|r|/\gamma'$ is not small enough}{
 $\rho=(r^*,r)$, $\beta= \frac{\rho \alpha'}{\rho'\omega'}$\\
 $p:= \beta ( p-\omega' v ) + r$ \mycomment{// probe vector}\\
 $v= Ap$ \\
 $\alpha = \frac{\rho}{(r^*, v)}$ \\
 $r:=r-\alpha v$ \\
 $t= Ar$ \\
 $\omega = \frac{(t,r)}{(t,t)}$ , $\hat{\omega} = \frac{|(t,r)|}{\sqrt{ (t,t) (r,r)}}$\\
 if($ \hat{\omega}  < \omega_0$) {
 $\omega:= \omega \omega_0/\hat{\omega}$
 }\\
 $x:= x+ \frac{\lambda'}{\gamma'} \omega r + \frac{\lambda'}{\gamma'}\alpha p$ \mycomment{// update the solution}\\
 $r:= r -\omega t$  \mycomment{// update the residual vector} \\
 $g=|r|$, $\gamma = \sigma /g$\\
 $r:= \gamma r$ \\
 $\gamma:= g/|r|$ \mycomment{// recalculate $\gamma$} \\
 $p:=\gamma p$, $v:=\gamma v$ \mycomment{// rescaling working vectors}\\
 $\omega'=\omega$, $\alpha'=\alpha$, $\rho'=\gamma \rho$\\
 $\gamma':=\gamma' \gamma$ \\
 $\lambda=\sigma_\lambda/|x|$ \\
 $x:= \lambda x$ \mycomment{// rescaling solution vector} \\
 $\lambda':= \lambda' \lambda$
}
\end{algorithm}

\section{Related Works}
\label{sec:Related_works}

A mixed precision acceleration is one of the standard techniques in linear solvers
for lattice QCD simulations.
It was demonstrated that a domain-decomposed preconditioner in FP32 with SIMD architecture
is efficient in Ref.~\cite{Luscher:2003qa}. 
The potential of using FP32 arithmetics in GPU for lattice QCD was shown by~\cite{Egri:2006zm}. 
Clark {\it et al.} \cite{CLARK20101517} tested a mixed precision solver with FP16
on NVIDIA GPUs, where they used a reliable update algorithm.
Their library, QUDA \cite{QUDA}, uses low-precision arithmetics intensively,
including FP16 and BF16.
Reference~\cite{Clark:2023xnn} gives a review of the mixed precision scheme in QUDA. 
Major modern lattice QCD code sets support mixed-precision solver with FP32.
Among them, GRID~\cite{Grid} and Bridge++~\cite{Aoyama:2023tyf, Bridge} have
tuning for the ARM architecture. QCD Wide SIMD library (QWS) \cite{Ishikawa:2021iqw}, designed for Supercomputer Fugaku, also has an implementation of preconditioner with FP16 
but its performance and effectiveness are not reported.

Apart from LQCD application, a mixed-precision BiCGStab with flying restart \cite{anciauxsedrakian:hal-05326901} has been recently proposed. 

\section{Implementation}
\label{sec:Implementation}

\begin{table*}[tb]
\caption{
Comparison of floating-point arithmetics,
bit length of each component, omitting the common 1-bit of sign,
the minimum and the maximum values of the exponent, and type in ACLE.
}
\begin{tabular}{crrrrl}
\hline
     & exponent & fraction &  $e_\text{min}$ & $e_\text{max}$ & \ \ type in ACLE \\
\hline
FP64 & 11 bit & 52 bit & $-2046$ & $2047$ & [\texttt{sv}]\texttt{float64\_t\ } \\
FP32 & ~8 bit & 23 bit & $-126$  & $127$  & [\texttt{sv}]\texttt{float32\_t\ } \\
FP16 & ~5 bit & 10 bit & $-14$   & $15$   & [\texttt{sv}]\texttt{float32\_t\ } \\
BF16 & ~8 bit & ~7 bit & $-126$  & $127$  & [\texttt{sv}]\texttt{bfloat32\_t} (SVE2) \\
\hline
\end{tabular}
\label{tab:floating-point}
\end{table*}

Let us first summarize the general properties of FP16 and the instruction sets
we can use for the implementation.
The IEEE 754 standard specifies floating point formats: FP64, FP32, and FP16. 
Table~\ref{tab:floating-point} summarizes the features of these floating-point
operations, except for BF16, which is an industry-defined format widely used
in hardware for machine learning.
The normalized FP16 numbers range from $\sim 6.10 \times 10^{-5}$ to $65504$.

The Scalable Vector Extension (SVE) by ARM is the first SIMD extension
for general-purpose CPUs to provide full support for half-precision arithmetics,
and the Fujitsu A64FX processor is the first Aarch64 processor to implement that extension.
In the initial SVE set, IEEE 754 binary16 format (FP16) was supported.
A later update in the SVE2 set added support for the BF16 format.
SVE2 is supported in NVIDIA Grace and Amazon Graviton3.
As the system we use is A64FX, we focus the implementation within the initial SVE set with FP16.
The C/C++ compiler by Fujitsu supports two kinds of FP16 implementation on A64FX. 
One is \texttt{\_\_fp16} data type, defined in the ARM C Language Extension (ACLE)
as IEEE 754-2008.
It is only for storing data and type conversion.
The arithmetic operations are performed by converting to floats.
This type could reduce the amount of memory access and B/F ratio so that the performance may improve, but we cannot use the feature of FP16 arithmetics supported by the CPU.
We therefore use the other data type, namely, \texttt{\_Float16}.
It is defined as an extension of the C11 standard as ISO/IEC TC 18661-3:2015.
The arithmetic operations with \texttt{\_Float16} are performed in FP16
so that one can expect a better performance than using \texttt{\_\_fp16}.
One complication of using \texttt{\_Float16} is that we need to write
an explicit type conversion to float or double in the code,
as the implicit casting is not supported.

For the SIMD implementation with FP16, we use the \texttt{svfloat16\_t} type in the ACLE.
Note that not all SIMD instructions support all three FP16/FP32/FP64 datatypes equally,
and some instructions do not support 16-bit datatypes.
Typical examples are the gather load and scatter store instructions, as well as
the horizontal \texttt{compact} instruction we used in \cite{HPC-Asia2023},
which support only 32- and 64-bit datatypes.
The compact operation collects the active elements, {\it i.e.},
true elements specified in the predicate, and packs them.
In practice, the compact operation is always followed by a store operation to memory.
We therefore first store the value of a \texttt{svfloat16\_t} type variable to an array of
\texttt{\_Float16} and then copy the contents of the array to the memory element-by-element
with an operation equivalent to the compact.

Even if the vectors are in FP16, we use FP32 for the summation within the OpenMP thread
and FP64 for reduction over the threads and global reduction with MPI.
To convert FP16 to FP32, we use \texttt{svrev\_f16} and \texttt{svcvt\_f32\_f16\_z}
as listed in Fig.~\ref{code:read_to_fp32}.
Datatype conversion for different element sizes in the SVE set targets only
the even-numbered lanes of the smaller element size.  That is, it converts only
the even-numbered elements in \texttt{svfloat16\_t}.
The \texttt{rev} instruction reverses the order of the elements, resulting in
odd-numbered elements coming to even-numbered lanes.
Finally, in \texttt{y1}, the promoted even-numbered elements of \texttt{x1} are stored,
and in \texttt{y2},
odd-numbered elements are stored in the reversed order.
Although the order of the elements is not preserved, it remains valid for reductions.
In Fig.~\ref{code:reduction}, we also list the pseudo code of the reduction.
To accumulate FP32 variables while maintaining accuracy, we combine partial summations
in a hierarchical manner.

\begin{figure}
\caption{
A code to read from an array of \texttt{\_Float16} to \texttt{svfloat32\_t}. 
This code does not preserve the order of the elements, but is valid for the reduction purpose.
}
\label{code:read_to_fp32}
\begin{lstlisting}
 // _Float16 *ptr is given
 svbool_t pg16 = svptrue_b16();
 svbool_t pg32 = svptrue_b32();
 svfloat16_t
   x1 = svld1_f16(pg16, (float16_t*)ptr);
 svfloat16_t x2 = svrev_f16(x1);
 svfloat32_t y1 = svcvt_f32_f16_z(pg32, x1);
 svfloat32_t y2 = svcvt_f32_f16_z(pg32, x2);
\end{lstlisting}

\end{figure}

\begin{figure}
\caption{Pseudo code to calculate norm squared.
Here, the site degrees of freedom are used for SIMD and thread parallelization,
and \texttt{site\_in} refers to inside the SIMD variable.
Each lattice site has \texttt{num\_on\_site} variables.
}
\label{code:reduction}
\begin{lstlisting}
// index of array :
// 32*(num_on_site*site_out + on_site) + site_in 
vx1=0, vx2=0   // FP32 vector varialbes
for site_out in the thread {
   vy1=0, vy2=0  // FP32 vector variables
   for on_site {
      // vz1, vz2 are FP32 vector variables
      cast and load to vz1, vz2 from \
      the array of _Float16
      vy1+=vz1*vz1, vy2+=vz2*vz2
   }
   vx1+=vy1, vx2+=vy2
}
vx1+=vx2
x = horizontal sum of vx1
sum x over threads and MPI procs in double prec.
\end{lstlisting}
\label{fig:norm}
\end{figure}

For the implementation, we use a general-purpose lattice QCD
code set Bridge++ \cite{Bridge, Ueda:2014rya} in this work.
It is written in C++ based on the object-oriented design,
and after its version 2.0, a tuned code for a specific architecture
is incorporated as an alternative to the base code.
For the A64FX architecture, such a tuned code called QXS branch
is available, which uses the ARM C Language Extension (ACLE)
to exploit the SIMD vector arithmetics and Fujitsu extension
of MPI \cite{Akahoshi:2021gvk, Aoyama:2023tyf}.
The tuning for the A64FX architecture in Bridge++ is found in
\cite{HPC-Asia2023} in some detail, including the implementation
using the ACLE intrinsics.
In Bridge++, the site degree of freedom in the $(x,y)$-plane is
packed into a SIMD vector.
Just like FP64 and FP32, the $x$-$y$ tiling size can be chosen to match
the SIMD vector size, which is 32 FP16 numbers, at compile time.
In this work, tile sizes of $4\times 2$, $4\times 4$, and $8\times 4$ are
used for FP64, FP32, and FP16, respectively.

\section{Results}
\label{sec:Results}

We use Supercomputer Fugaku at RIKEN Center for Computational Science for
numerical investigation.
The Fujitsu C/C++ compiler with the language environment lang/tcsds-1.2.42 is used
with the compile options
\texttt{-Nclang -fopenmp -Ofast -mllvm -inline\textbackslash\\-threshold=1000}.
The option \texttt{-Ofast}
derives \texttt{-Kfz},
which flashes subnormal floating-point numbers to zero.
We follow the default CPU mode of the system as of November 2025, called boost eco mode;
the frequency is 2.2 GHz and one of the two arithmetic pipelines is stopped.
The theoretical peak performance of this mode is 1690 GFlops per CPU.
We use 64 MPI processes on 16 nodes.

The lattice size is set to $32^3\times 64$.
For the gauge configuration $\{U_\mu(x)\}$, which determines the Wilson matrix $D_W$,
the standard test configuration in the Bridge++ code sets \cite{Bridge} is used%
\footnote{
Comment for reproducibility:
After reading the configuration from the file,
three iterations of the stout smearing (with $\rho=0.1$) are applied to make the configuration smoother and closer to the practical situation.
},
whose size is $4^3\times 8$ and periodically extended to any lattice size.
The hopping parameter in the Wilson matrix is set to $\kappa=0.13$.
Just for reference, the ratio of the largest and smallest eigenvalues
in the absolute value is $1.22/0.069=17.7$.

\begin{figure}
\noindent\centering
\includegraphics[width=\linewidth]{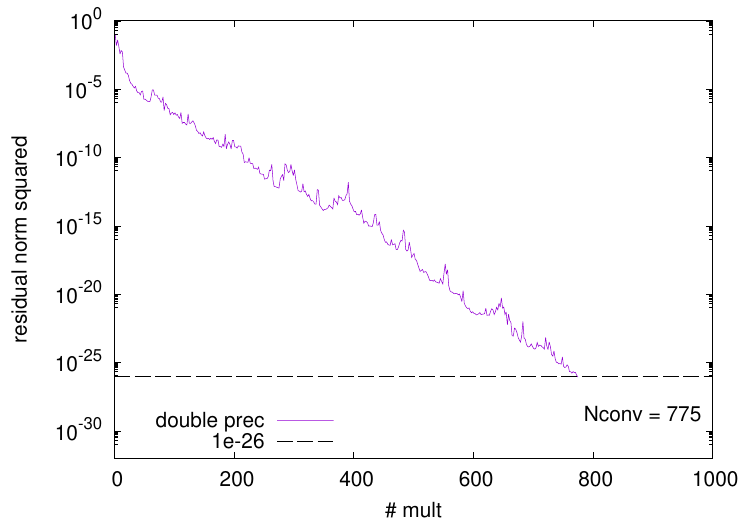}
\caption{
Convergence of the FP64 BiCGStab solver:
Residual norm squared versus number of matrix-vector multiplications.
}
\label{fig:baseline_FP64}
\end{figure}

\begin{figure}
\noindent\centering
\includegraphics[width=\linewidth]{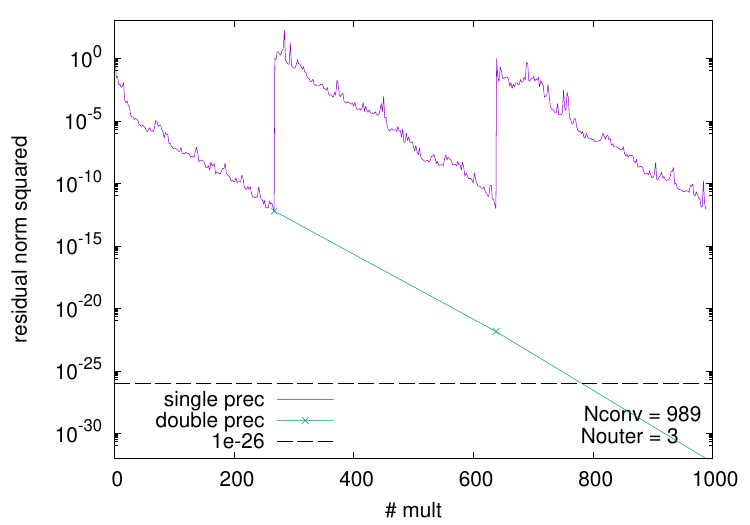}
\includegraphics[width=\linewidth]{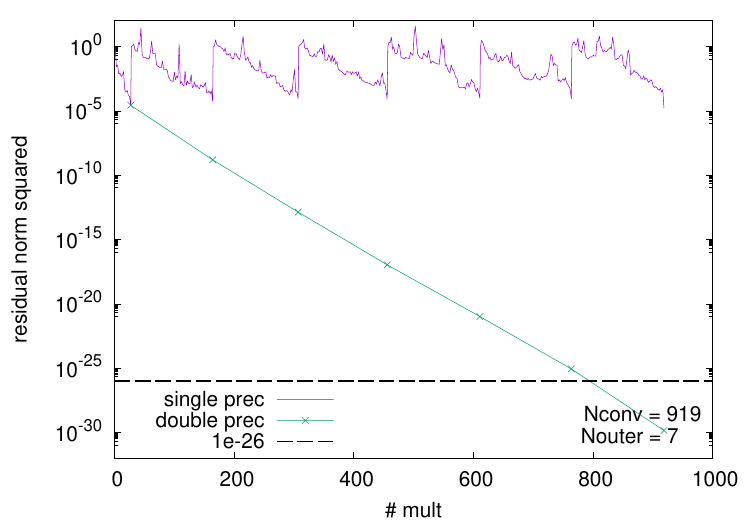}
\caption{
Convergence of the FP32/FP64 mixed precision solvers.
Residual norm squared versus number of matrix-vector multiplications,
for the inner solver's convergence criteria set at $10^{-12}$ (top panel) 
and $10^{-4}$ (bottom). 
The purple and green curves indicate the residual norms
of the inner low-precision and outer FP64 solvers, respectively.
}
\label{fig:baseline_FP32}
\end{figure}

Let us start with applying the original BiCGStab solver in FP64 precision,
as shown in Figure~\ref{fig:baseline_FP64}.
We set the known vector as a unit vector, $b=(1,0,0,\dots,0)$ throughout
this work.
The convergence criterion is set to $10^{-26}$ for the squared norm of the
residual vector.
As the figure exhibits, the BiCGStab solver with FP64 successfully applies
to the current problem, showing a typical convergence behavior of this algorithm.

The mixed precision solver with Algorithm~\ref{alg:mixed_org}
combined with FP32 preconditioning with Algorithm~\ref{alg:bicgstab_rescaled}
(without rescaling: $\gamma=1$) is applied to the same problem in
Figure~\ref{fig:baseline_FP32}.
The criterion for the low-precision solver is $10^{-12}$ (top panel)
and $10^{-4}$ (bottom).
The latter condition is applied for comparison with the FP16 preconditioner
examined in the following.
As displayed in Table~\ref{tab:etime}, the mixed precision solver
successfully reduces the elapsed time for convergence for
both the inner BiCGStab convergence conditions.

\begin{table*}[tb]
\caption{
Elapsed time to solve the full system and the number of matrix-vector multiplications
for the low-precision solvers in the mixed-precision case under various setups.
The target tolerance squared, ${\epsilon_\text{tol}}^2$, is for the low-precision solvers
with FP32 and FP16.
}
\centering
\begin{tabular}{ccccccc}
    \hline
      scheme   &  $s$  & $\sigma$ & $\sigma_\lambda$ & low prec.\ ${\epsilon_\text{tol}}^2$ &
      \#mult (matrix-vector) & elapsed time [sec.] \\
    \hline 
      double          & --- & --- & --- & --- & 775 & 1.39 \\
    \hline
      mixed with FP32 &  1  & --- & --- & $10^{-12}$ & 989 & 0.96 \\
      mixed with FP32 &  1  & --- & --- & $10^{-4}$ & 919 & 0.92 \\
    \hline
      mixed with FP16 &  128 & 64 & --- & $10^{-4}$ & 858 & 0.46 \\
                      & 4096 & --- & --- & $10^{-4}$ & 920 & 0.47 \\
                      &  128 & 256 & 256 & $10^{-4}$ & 834 & 0.46 \\
    \hline                  
\end{tabular}
\label{tab:etime}
\end{table*}

\begin{figure}
\noindent\centering
\includegraphics[width=\linewidth]{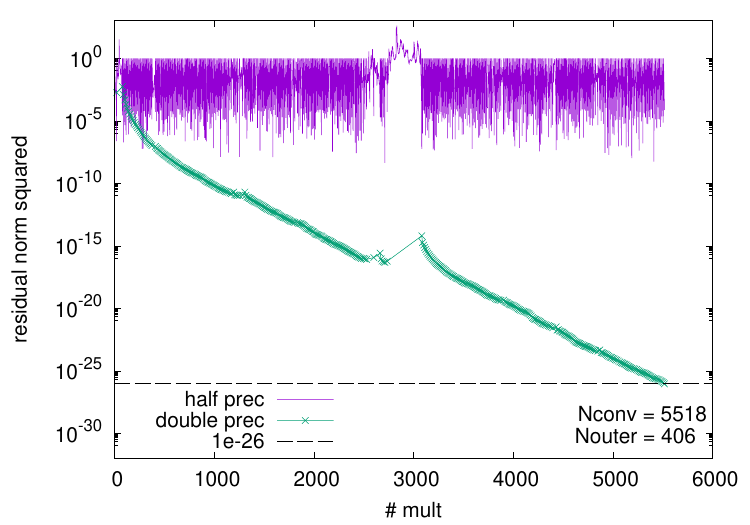}
\caption{
Convergence of FP16/FP64 mixed precision solver without rescaling:
Residual norm squared versus number of matrix vector multiplications.
The purple and green curves indicate the residual norms
of the inner low-precision and outer FP64 solvers, respectively.
}
\label{fig:baseline_FP16}
\end{figure}

Figure~\ref{fig:baseline_FP16} shows the mixed precision solver with FP16.
The convergence criterion for the low precision solver is $10^{-4}$.
The solver eventually converges, but its convergence is very slow.
The low-precision solver once suffers from stagflation,
which causes a large increase in the total residual norm.
This is considered due to the unstable convergence of the FP16 BiCGStab solver.

\begin{figure}
 \noindent\centering
 \includegraphics[width=\linewidth]{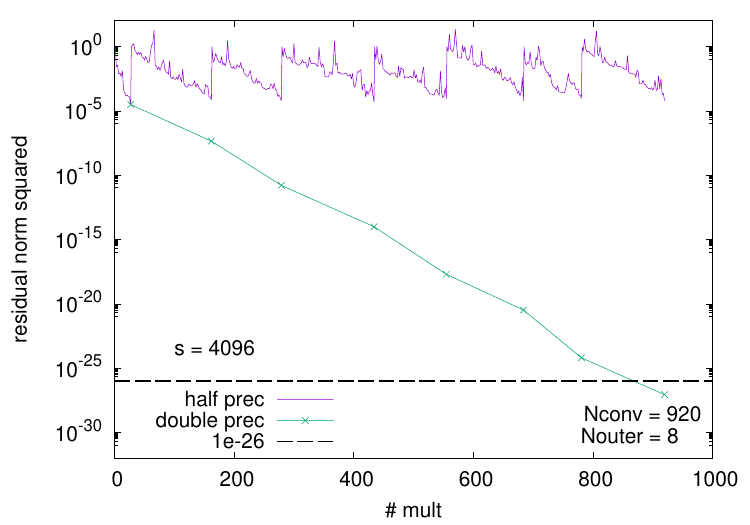}
 \includegraphics[width=\linewidth]{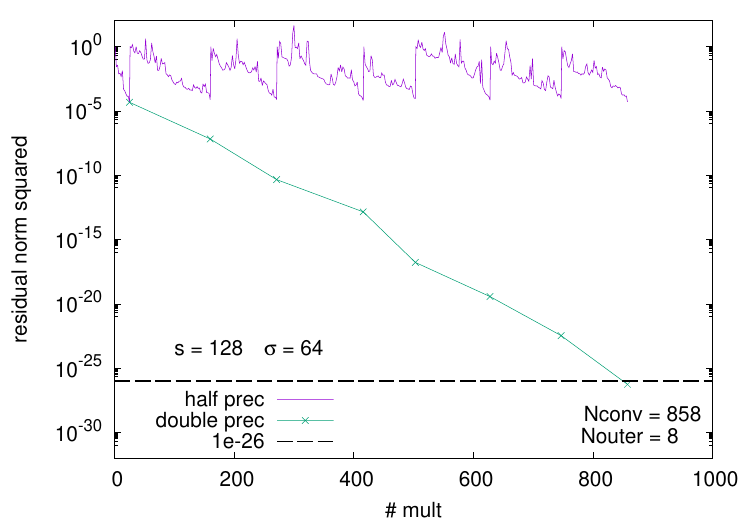}
\vspace{-0.2cm}
\caption{
Convergence of FP16/FP64 mixed precision solver with rescaling:
Residual norm squared versus number of matrix vector multiplications.
The purple and green curves indicate the residual norms
of the inner low-precision and outer FP64 solvers, respectively.
In the top 
panel, only the rescaling in the Richardson algorithm is applied with
the normalization factor $s=4096$.
In the bottom 
panel, rescaling of the residual vector in the inner BiCGStab solver
is also applied with $s=128$ and $\sigma=64$.
}
\label{fig:rescaled_FP16}
\end{figure}

We then apply the proposed method in Section~\ref{sec:Algorithm}, {\it i.e.},
a combination of Algorithms~\ref{alg:mixed} and \ref{alg:bicgstab_rescaled},
or a combination of \ref{alg:mixed} and \ref{alg:bicgstab_rescaled2}.
In Figure~\ref{fig:rescaled_FP16}, we plot the results with solely rescaling the input vector
(top panel) and rescaling both input and residual vectors (bottom).
The convergence behavior with FP16 is drastically improved.
For comparison, plotted in the bottom panel of Figure~\ref{fig:baseline_FP32}
is a mixed precision result with FP32, where the convergence criterion for
the low-precision solver is the same as in FP16 cases. 
Although the outer iterations slightly increase in FP16 cases, the total number of
matrix-vector multiplications is almost the same or even fewer than in the FP32 case.

\begin{figure*}
\centering
\includegraphics[width=0.499\linewidth]{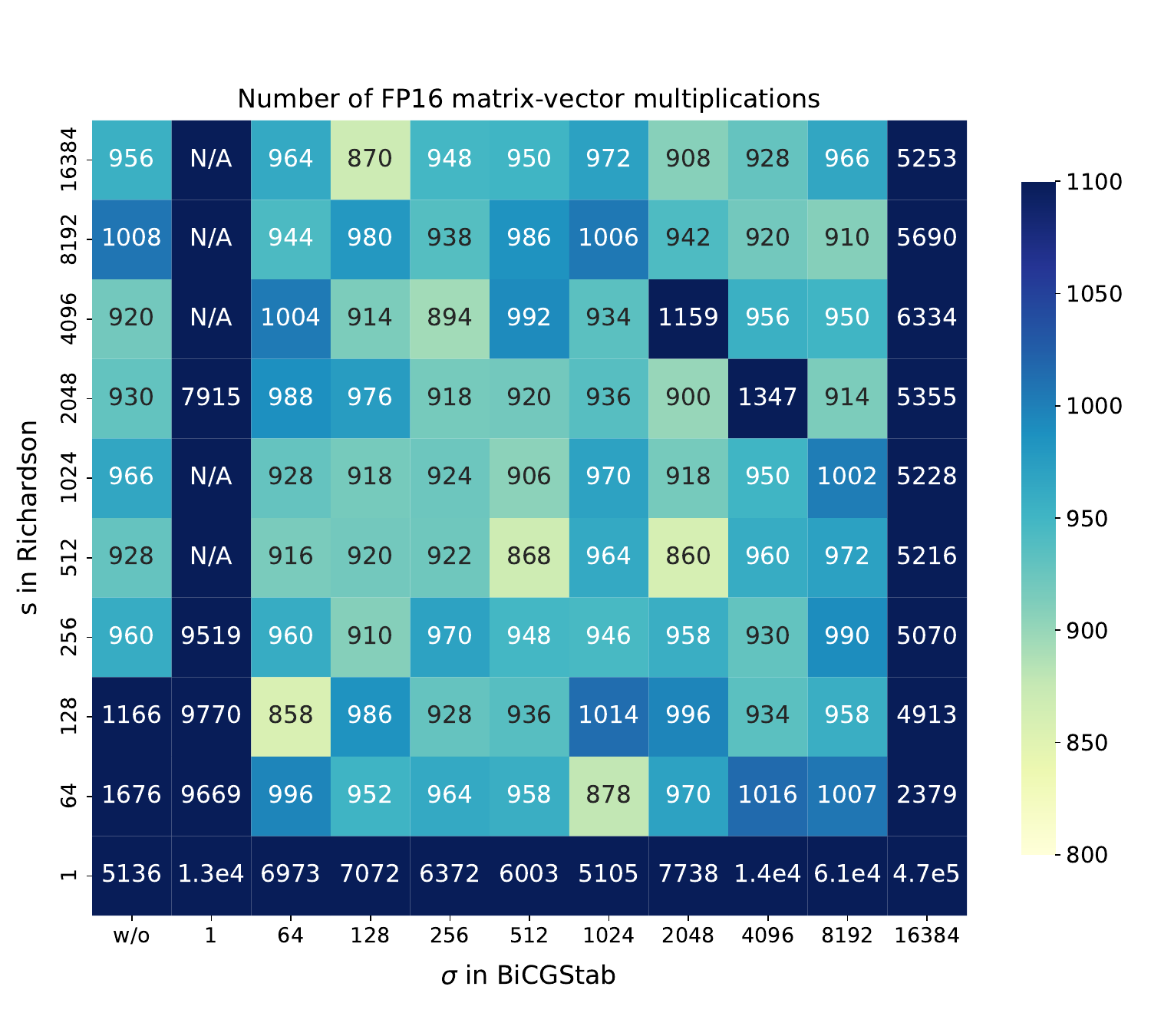}%
\includegraphics[width=0.499\linewidth]{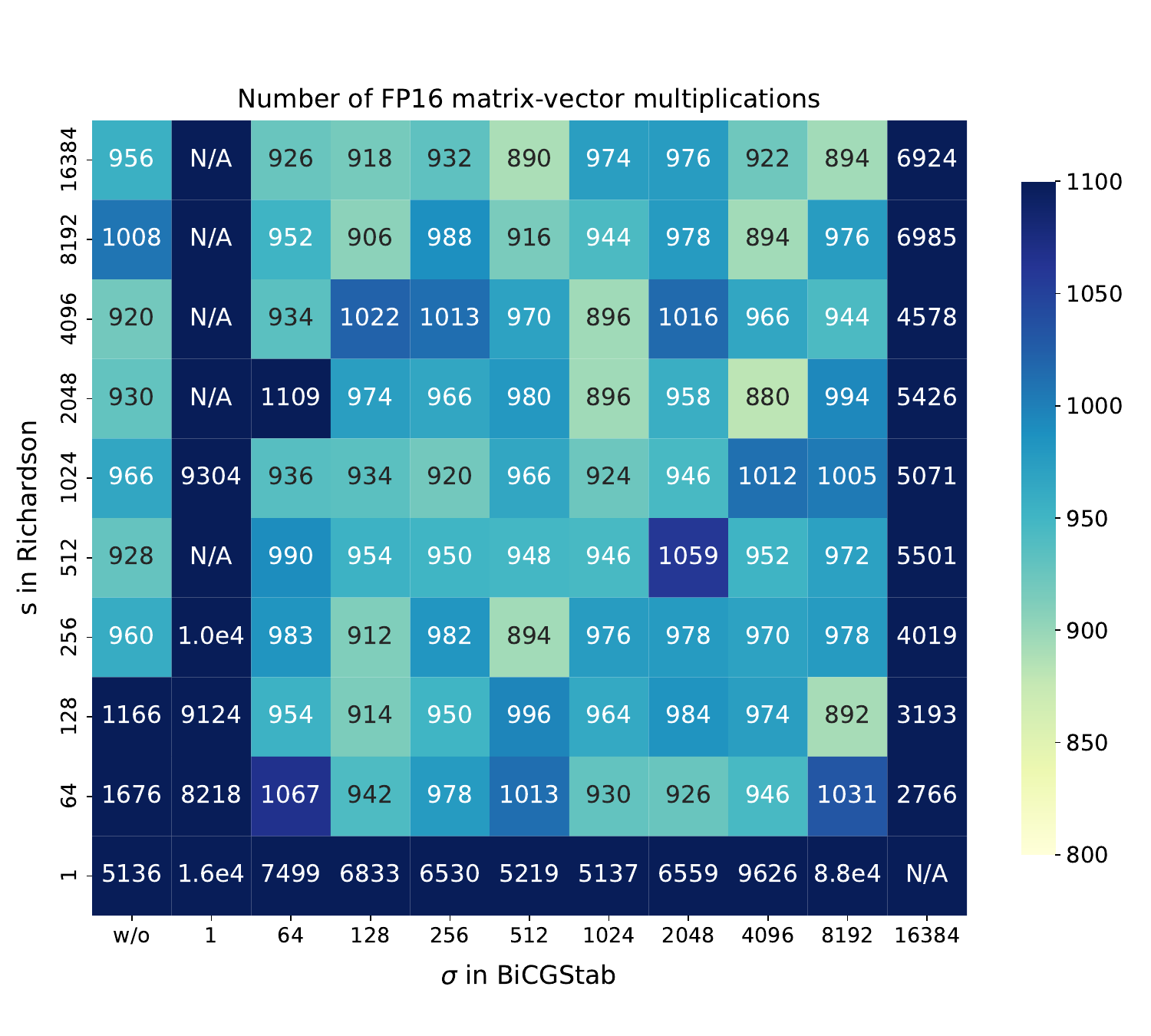}
\vspace{-0.8cm}\\
\includegraphics[width=0.499\linewidth]{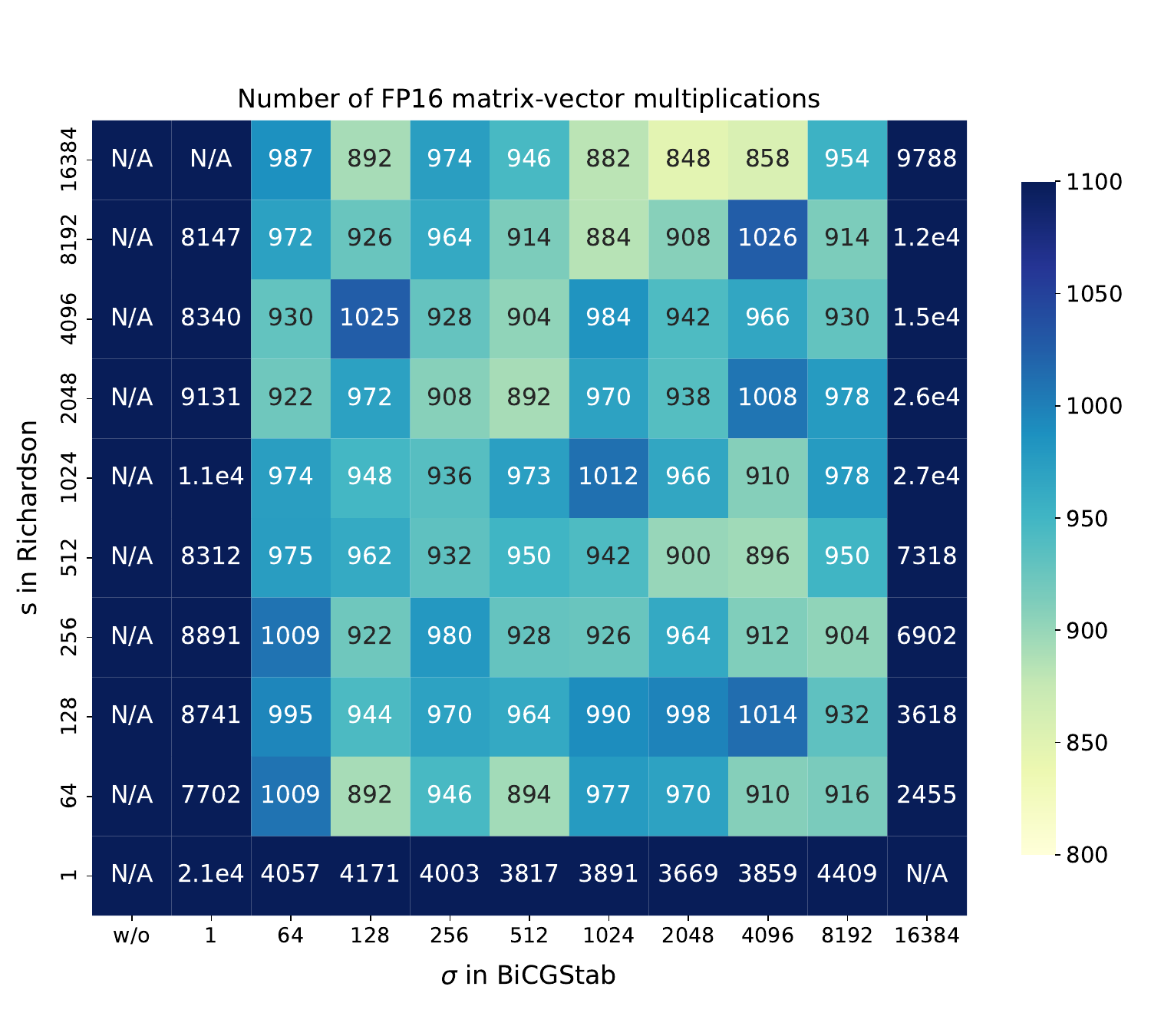}%
\includegraphics[width=0.499\linewidth]{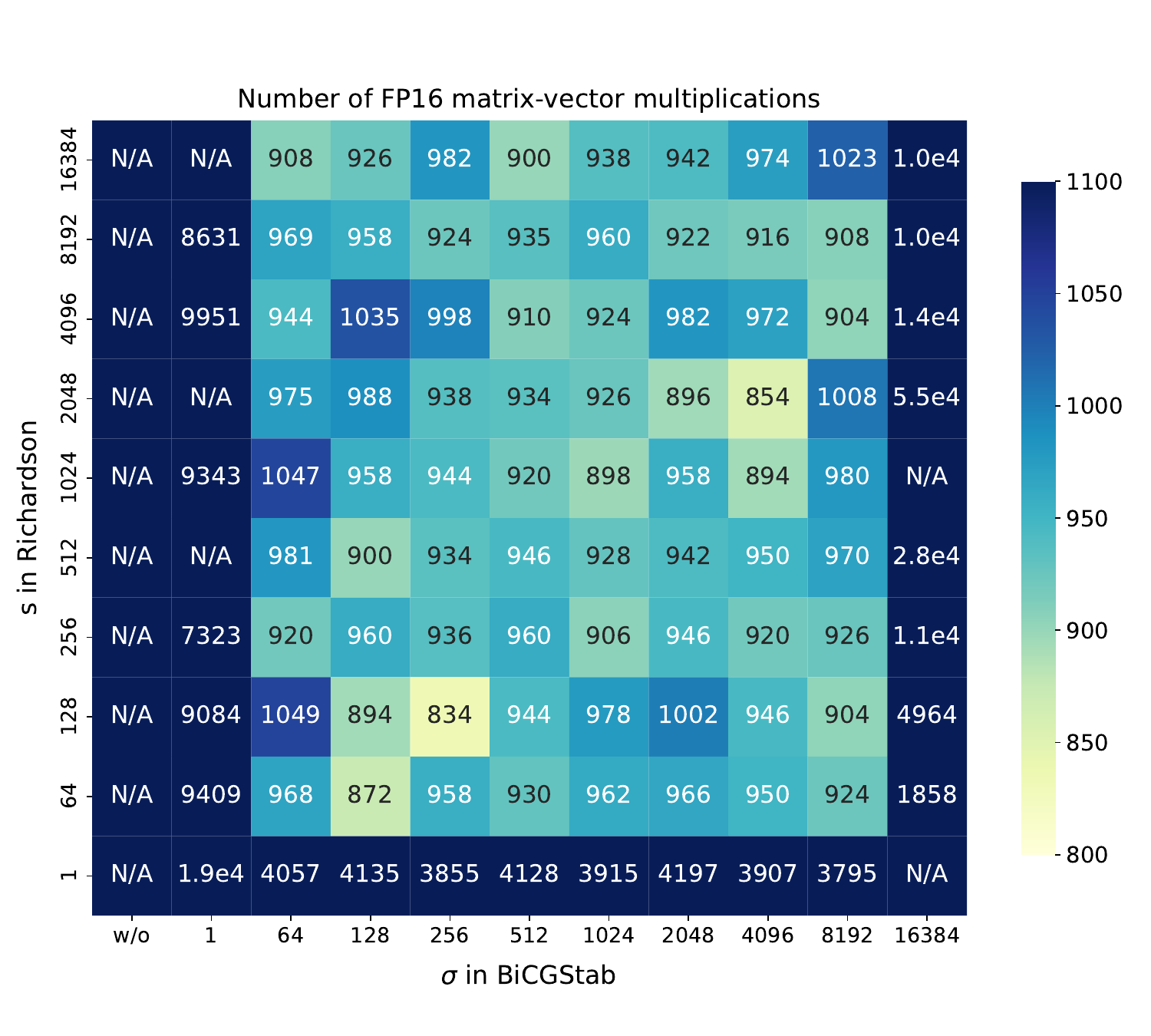}
\caption{
Tableaus to represent the dependence of the number of FP16 matrix-vector multiplications
on the rescaling parameter $s$ in the iterative refinement (vertical axis) and
rescaling parameter $\sigma$ of residual norm in BiCGStab (horizontal axis). 
The rescaling of the solution is off (Alg.~\ref{alg:bicgstab_rescaled}) in the top panels
and on (Alg.~\ref{alg:bicgstab_rescaled2}) in the bottom panels.
In each case, recalculating $\gamma$ is off in the left panel, and on in the right panel.
To summarize, (rescaling solution, recalculating $\gamma$) = (off,off): top-left,
(off,on): top-right, (on,off): bottom-left, (on,on): bottom-right.
The normalization of the solution vector $\sigma_\lambda$ in Alg.~\ref{alg:bicgstab_rescaled2}
is set to the same as the normalization of the residual vector, $\sigma$.
Because of this choice, we do not have results without rescaling the residual norms
for the bottom panels (N/A).
The other (N/A)s represent non-converging results.
}
\label{fig:count_2d}
\end{figure*}

Now we examine the optimal values for the parameters $\sigma$ (in BiCGStab) and
$s$ (in Richardson) concerning the number of FP16 matrix-vector multiplications.
Figure~\ref{fig:count_2d} shows the number of low-precision matrix-vector multiplications
for various cases.
The rescaling of the solution is off (Alg.~\ref{alg:bicgstab_rescaled}) in the top panels
and on (Alg.~\ref{alg:bicgstab_rescaled2}) in the bottom panels.
In each case, recalculating $\gamma$ is off in the left panel, and on in the right panel.
The normalization of the solution vector $\sigma_\lambda$ in Alg.~\ref{alg:bicgstab_rescaled2}
is set to the same as the normalization of the residual vector, $\sigma$.
Although the bottom-left corners in the upper panels are the settings closest to
the unimproved case in Fig~\ref{fig:baseline_FP16}, the iterative refinement with
the recalculated $\hat{s}$ slightly reduces the count from 5518 to 5136.
Except for the setting $\sigma=1$ or $s=1$, for a wide range of choices
of $\sigma$ and $s$, we observe a significant improvement with the
iteration counts $\sim 850\text{--}1000$.
We also observe that there is no clear preference among the four panels;
whether the rescaling factor $\gamma$ is recalculated or not, or whether the
solution vector is rescaled or not, does not affect the convergence behavior.
The fluctuation due to the choice of the parameters is rather large.
Since our target problem, lattice QCD, the matrix norm of the fermion matrix
$D_W$ is constrained below a certain value, we can set the norm of the input vector,
$s$, close to the maximum allowed number in FP16 without having overflow.

\begin{figure}[thb]
\centering
\includegraphics[width=\linewidth]{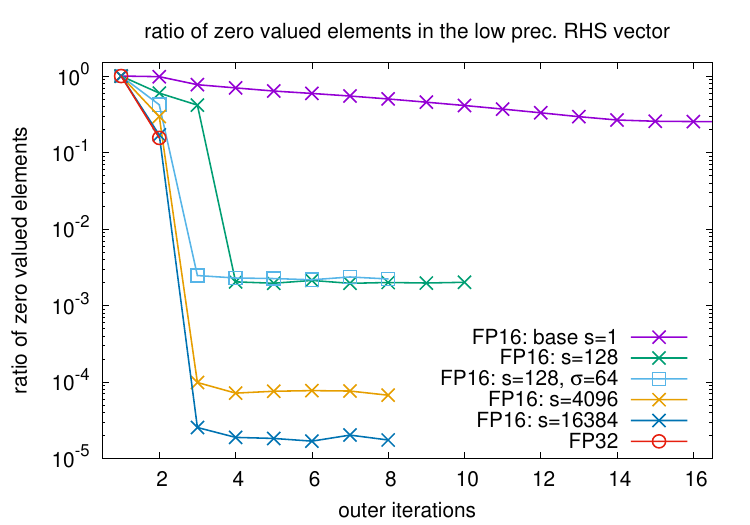}
\caption{
Fraction of the zero-valued elements in the input vector of the low-precision solver.
The rescaling of the residual vector is applied only when $\sigma$ is specified in the legend.
After the second outer iteration, there is no zero-valued element in the FP32 case.
}
\label{fig:zeros}
\end{figure}

We can confirm that the underflow has been largely removed after the improvement.
Figure~\ref{fig:zeros} shows the number of zero-valued elements
in the input vector of the low-precision solvers.
In the first iteration, only one element has a non-zero value out of 25,165,824 elements.
The red circle in the plot is the FP32 case with the same convergence condition as FP16,
and non-zero elements disappear after the second iteration.
Roughly speaking, after the second iteration the value represents
the fraction of the underflowed elements.
As the normalization for the input vector becomes larger, 
we have less underflow as expected.
The iterative steps can be regarded as a diffusion process, and having non-zero values
implies that the information of the original input vector has propagated to that element.
The comparison in $s=128$ cases implies that having the residual vector rescaled
($\sigma=64$) propagates information of the input more efficiently. 

We finally compare the elapsed time and performance.
Table~\ref{tab:etime} summarizes the elapsed time of the full solver for selected cases,
together with double precision and mixed-precision with FP32 cases.
The elapsed time in the FP16 cases is about half of that in the FP32 cases
and about 1/3 of the FP64 case.
The performance of the matrix-vector multiplication is observed as
2045 GFlops, 3895 GFlops, and 8249 GFlops for FP64, FP32, and FP16 cases, respectively.

\section{Conclusions and Outlooks}

We applied half-precision floating-point numbers (FP16) in mixed-precision solvers
for lattice QCD simulation on Supercomputer Fugaku.
A naive extension of a mixed-precision solver with FP32 was numerically unstable
due to underflow in FP16.
We introduced rescaling in the iterative refinement and low-precision BiCGStab solver
to bypass the underflow, and achieved a two-times of speed up compared to the FP32 case.
Although we examined only the BiCGStab solver, the proposed rescaling methods should be
applicable to other iterative solver algorithms.

In this work, we used a simple Wilson fermion matrix.
The implementation of matrix-vector multiplication makes use of the structure
of the matrix and consists of several internal steps.
In practical lattice QCD simulations, more complex fermion matrices such as 
the clover and domain-wall types are used.
It is important to study these matrices as they have more complicated internal
steps and have more potential for rounding errors. 
Finally, it would be interesting to compare results with mixed-precision using BF16.
Since the narrow dynamic range of FP16 is compensated for by rescaling in
the proposed method, the difference in the size of the fraction bits may result in
significant differences in the iteration counts.
It is also important to implement the code for the GPU, in particular,
exploiting the Tensor Core.

\begin{acks}
I.~K. thanks Atsushi~Suzuki in R-CCS
for useful comments.
Numerical computations are performed on Supercomputer Fugaku at RIKEN Center for
Computational Science (ra250006).
Some of the code developments were performed on Wisteria-O at the University of Tokyo
through the Multidisciplinary Cooperative Research Program in CCS, University of Tsukuba.
The calculation of the condition number was performed on Miyabi-G through
the same program in CSS, University of Tsukuba.
This work is supported by JSPS KAKENHI (%
JP22H01224,
JP25H01109).
\end{acks}

\bibliographystyle{ACM-Reference-Format}
\bibliography{hpcasia}


\end{document}